\documentclass[journal]{IEEEtran}

\usepackage{amsmath}
\usepackage[tight]{subfigure}
\usepackage[footnotesize]{caption} 
\usepackage{cite}
\usepackage{eurosym}
\usepackage{amsfonts}
\usepackage{graphics}
\usepackage{epsfig}
\usepackage{makecell}
\usepackage{psfrag}
\usepackage{pstricks}
\usepackage{array}
\usepackage{shortvrb}
\usepackage{epsf}
\usepackage{graphicx}
\usepackage{rotating}
\usepackage{float}
\usepackage{color}
\usepackage{cite}
\usepackage{soul}
\usepackage{xcolor}
\usepackage{multirow}
\usepackage{comment}
\usepackage{acronym}
\usepackage{url}
\usepackage{breakurl}
\usepackage[breaklinks, hidelinks]{hyperref}

\IEEEoverridecommandlockouts
\usepackage[utf8]{inputenc}
\begin{document}

\title{Day-Ahead Forecasting of Largest Single Infeed/Outfeed  on the Irish Power Grid: A Generative Artificial Intelligence Approach}

\author{ \IEEEauthorblockN{ Amir Moshari,\IEEEauthorrefmark{1} Mo  Cloonan,\IEEEauthorrefmark{1}  Taulant K\"{e}r\c{c}i,\IEEEauthorrefmark{1}
    \IEEEmembership{Senior Member,~IEEE}, Zhi Li,\IEEEauthorrefmark{2} Colm Gaffney,\IEEEauthorrefmark{1} \\ Chotiya Mahittigul,\IEEEauthorrefmark{2}  Manuel Hurtado,\IEEEauthorrefmark{1} \IEEEmembership{Member~IEEE}, Simon Tweed,\IEEEauthorrefmark{1} \\ Bryan Murray,\IEEEauthorrefmark{1} Michael Walsh,\IEEEauthorrefmark{2}  Eoin Kennedy,\IEEEauthorrefmark{1} and Ritesh Madan\IEEEauthorrefmark{2}
    }\vspace*{0.3cm}
  \IEEEauthorblockA{
    \begin{tabular}{cc}
      \begin{tabular}{@{}c@{}}
        \IEEEauthorrefmark{1}
        Transmission System Operator \\
        Ireland
      \end{tabular} &
      \hspace{0.3cm}
      \begin{tabular}{@{}c@{}}
        \IEEEauthorrefmark{2}
        GridZero.ai \\Ireland \& USA
      \end{tabular} 
    \end{tabular}
  }
}

\IEEEoverridecommandlockouts

\maketitle
\IEEEpubidadjcol

\begin{abstract}
  
  This paper presents a generative artificial intelligence (Gen AI) approach for forecasting, at a day-ahead stage, the largest single infeed (LSI) and largest single outfeed (LSO) on the Irish power system to assist in reserve dimensioning.  Developed collaboratively between EirGrid, the electric transmission system operator (TSO) for Ireland,  and GridZero.ai using the GridZero.ai platform, the system delivers accurate forecasts up to 38 hours ahead of real-time using limited data available before the day-ahead and intra-day energy market gate closure timings.  Initial performance demonstrates an accuracy with a mean absolute percentage error (MAPE) that is only 1.1\% higher than the results possible using full market data (8-hours ahead).  Thus, if this approach is integrated into operational systems and such high levels of accuracy are maintained, reserve procurement costs could be significantly reduced. The results also demonstrate the practicality and extensibility of AI-powered resource planning for TSOs.

\end{abstract}


\section{Introduction}

\subsection{Motivation}

The All-Island power system (AIPS) of Ireland is known worldwide for integrating pioneering levels of variable non-synchronous renewable energy sources (RES) like wind and solar power.  Currently, EirGrid and SONI, the transmission system operators (TSOs) of Ireland and Northern Ireland, respectively, are able to safely and securely operate the AIPS with up to 75\% of instantaneous system non-synchronous penetration (SNSP) \cite{10253224}.  The operation of the AIPS at high SNSP levels requires careful consideration in terms of stability management, particularly frequency stability, given the relatively low-inertia levels and the small size of the island AC power system compared to the largest single infeed (LSI) or largest single outfeed (LSO) \cite{10688904}.  Specifically, the AIPS has a peak demand of approximately 7.5 GW compared to current LSI and LSO figures, which are around 500 MW and, in the near future, 700 MW with the connection of a new high voltage DC (HVDC) interconnector to France (i.e., LSI/LSO could account for around 10\% of peak demand) \cite{kerci2025comprehensive}.

\begin{figure}[!h]
  \begin{center}
    \resizebox{\linewidth}{!}{\includegraphics[trim={0 0 0 1.2cm},clip]{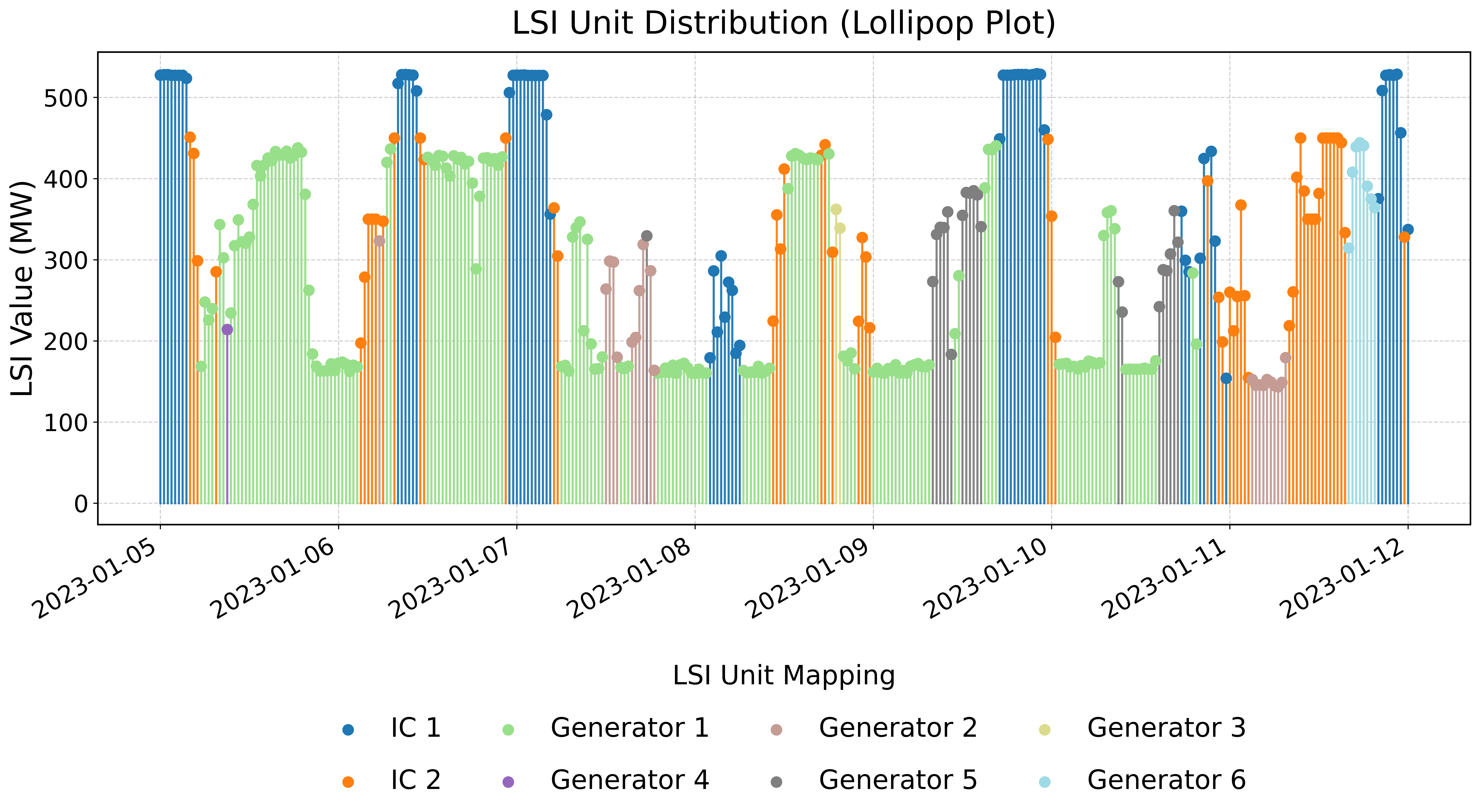}}
    \caption{Illustration of the variation of LSI and frequent LSI unit changes (represented by different colours) in the AIPS. LSO exhibits comparable volatility, though its occurrences are sparser since only interconnector exports contribute to LSO.}
    \label{fig:lsi_lollipop}
  \end{center}
  \vspace*{-4mm}
  \end{figure}

As part of a new day-ahead system services auction (DASSA) arrangement in Ireland and Northern Ireland that is expected to go live in May 2027, the TSOs, EirGrid and SONI, will procure reserve capacities at a day-ahead stage \cite{dassadesign}.  A key input to the dimensioning of these reserve requirements that will be procured in DASSA will be the forecasted LSI and LSO (see Section~\ref{sec:model} below) for each trading period (i.e., 30 minutes) \cite{dassavfm}.  Thus, the greater the accuracy of LSI and LSO forecasts, the greater the potential for lower reserve capacity procurement costs while maintaining system reliability.  However, the LSI and LSO vary significantly over the course of a day.  For instance, Figure~\ref{fig:lsi_lollipop} shows the variation of LSI over a week in January 2023 in the AIPS.  This paper presents a new generative artificial intelligence (Gen AI) approach to tackle the LSI/LSO forecasting problem.

\subsection{Literature Review}

The increasing variability introduced by RES and interconnector exchanges has driven extensive research on how TSOs dimension operating reserves.  Early work, such as \cite{DEVOS2019272} proposed dynamic reserve sizing frameworks that adjust with system conditions rather than fixed margins. In \cite{ROALD2023108725}, the authors review stochastic, robust, and chance-constrained optimization techniques for operation planning under uncertainty.  More recently, \cite{9744438} introduced probabilistic forecasts directly into reserve dimensioning, linking forecast accuracy to reserve efficiency.  Building on these ideas, \cite{10909649} and related multi-area studies \cite{KHODADADI2025111807} compared dynamic reserve and stochastic formulations using probabilistic inputs across interconnected systems, confirming that forecast-driven reserve management can improve reliability and reduce cost.

While prior studies have advanced optimization based approaches to reserve sizing once uncertainty is characterized, they typically assume the availability of high-quality probabilistic forecasts at operational timescales. Few explicitly address the challenge of forecasting the underlying contingencies, namely the LSI and LSO, at day-ahead horizons when such information is not yet available. In addition, most implementations remain at the simulation or proof of concept stage, leaving open how these methods would perform when integrated into TSO environments.

Reference \cite{vaswani2017attention} proposes a deep learning Transformer model architecture for modeling sequences where both short and long-term dependencies are captured in a computationally scalable manner.  While Transformer model architectures have revolutionized language models (see \cite{rothman2022transformers} for example), transformer models are also powerful in modelling time series (see \cite{wen2022transformers} for a survey). Compared to classical machine learning (ML), these models are more powerful in modelling spatiotemporal effects, and they adapt and generalize on different parts of the data distribution at a faster rate than classical ML. While they have been applied to certain forecasting problems (e.g., wind power \cite{li2024adaptive} and weather forecasting \cite{kurth2023fourcastnet}), our paper presents a novel application of  a Gen AI approach to decision making in the electricity grid.  We believe this approach should be utilised to support many other aspects of operational planning, grid optimization, and real-time management, especially as the integration of RES and storage assets make electricity system operation more complex and dynamic than previous.

\subsection{Contributions}
In summary, this paper brings the following main novel contributions.

\begin{itemize}
    \item Demonstration of a Gen AI approach that forecasts the LSI and LSO up to 38 hours ahead, supporting earlier reserve dimensioning in a real-world low-inertia power system, namely the AIPS.
    \item Introduction of an adaptive process to inform the integration of Gen AI into power system operations through a unified AI platform and a DevOps-based collaborative workflow.
    \item Validation of forecast performance within 1.1\% of the operational 8-hour rolling model based on full market data, indicating around 15\% potential reserve cost savings and extensibility to other planning tasks.
\end{itemize}

\subsection{Organization}

The remainder of the paper is organized as follows.  Section~\ref{sec:model} further discusses the need for LSI/LSO forecasting and presents the proposed Gen AI approach to this problem.  Section~\ref{sec:case} presents the main results of the paper.  Finally, conclusions and future work are discussed in Section~\ref{sec:conclu}.

\section{Problem to Solve and Methodology}
\label{sec:model}

\subsection{Need for Day-Ahead LSI and LSO Forecasting}

Recently, EirGrid and SONI completed a review of the reserve services to be procured in DASSA starting from May 2027 \cite{11225021}.  Table~\ref{tab:newreserve} summarizes these services, namely Fast Frequency Response (FFR), Primary
Operating Reserve (POR), Secondary Operating Reserve (SOR),
Tertiary Operating Reserves (TOR1 and TOR2), and Replacement
Reserve (RR).  In particular, the TSOs decided to introduce downward reserve services and procure FFR in three different sub-categories based on full activation time (FAT).  

The amount of reserves required to respond to events on the power system is directly related to the magnitude of the largest single contingency event.  For reserves planning on the AIPS this is the loss of LSI and LSO.  In this context, Table~\ref{tab:newreserve} provides indicative minimum volume requirements for each reserve service as a percentage of the reference incident (RI).  According to the European Union guideline on electricity transmission system operation \cite{sogl} ``\textit{the reference incident is the largest imbalance that may result from an instantaneous change of active power such as that of a single generating module, single demand facility or a single HVDC interconnector, and shall be determined separately for positive and negative directions''}.

\begin{table}[h!]
  \centering
  \caption{Summary of the proposed new DASSA reserve services.} 
  \label{tab:newreserve}
  \resizebox{1.0\linewidth}{!}{
  \begin{tabular}{cccccccc}
    \hline
    Reserve Service & Direction & Minimum Volume Requirement & FAT & Response Duration  \\
    & (Upward/Downward) & (\% of Reference Incident) & (s) &  \\
    \hline
    FFR (Cat. 1)& Upward \& Downward & 60-80 & 0.15 & Up to 10s  \\
    FFR (Cat. 2) & Upward \& Downward & 60-80 & $\leq$ 0.3 & Up to 10s  \\
    FFR (Cat. 3) & Upward \& Downward & 60-80 & $\leq$1 & Up to 10s \\
    POR & Upward \& Downward & 100 & 5 & Up to 15s   \\
    SOR & Upward \& Downward & 100 & 15 & Up to 90s \\
    TOR1 & Upward \& Downward & 100 & 90 & Up to 300s\\
    TOR2 & Upward \& Downward & 100 & 300 & Up to 1200s  \\
    RR & Upward \& Downward & 100 & 1200 & Up to 3600s \\
    \hline
  \end{tabular}}
\end{table}

The positive/negative RI is the sum of LSI/LSO and consequential losses ($\rm C_{\rm loss}$) triggered by the same incident:
\vspace{-1mm}
\begin{align}
  \label{eq:dassa} \rm RI^{-} &= 
  \rm LSO + C_{\rm loss} \, , \\
  \label{eq:dassa1} \rm RI^{+} &= 
  \rm LSI + C_{\rm loss} \, .
\end{align}

Note that $\rm C_{\rm loss}$ are typically inadvertent 
and caused by, for example, lack of fault-ride-through capability of the concerned demand/generation.  The TSOs will set the $\rm C_{\rm loss}$ component of the RI as a fixed percentage of the concerned generation and demand (e.g., 50\% of large energy user demand assumed to reduce automatically following a transmission fault). However, that is not the case for LSI and LSO, where the TSOs are expected to forecast these at a day-ahead stage, at 9 am.  

However, as shown in Figure~\ref{fig:lsi_lollipop}, LSI ( LSO is similar) varies significantly over the course of a day.  This is because relatively small changes in system conditions can drive significant changes in the output levels of individual interconnectors and generators.  EirGrid wishes to develop reliable forecasts at 9 am at the day-ahead stage to facilitate more efficient procurement of reserves.  This timing is in advance of the wholesale day-ahead electricity market, and two subsequent cross-border intra-day trading market windows, so market results and the market traded positions of generators (that drive interconnector flows) are not yet known.  
Traditional forecasting approaches struggle to provide reliable results at this day-ahead stage due to the large number of variables with high levels of uncertainty and the strong degree of interaction between them.  These approaches are based on numerical unit commitment style algorithms using a single forecast for key inputs such as RES and demand.

\subsection{AI Approach and Platform}

\begin{figure}[!h]
  \begin{center}
    \resizebox{\linewidth}{!}{\includegraphics{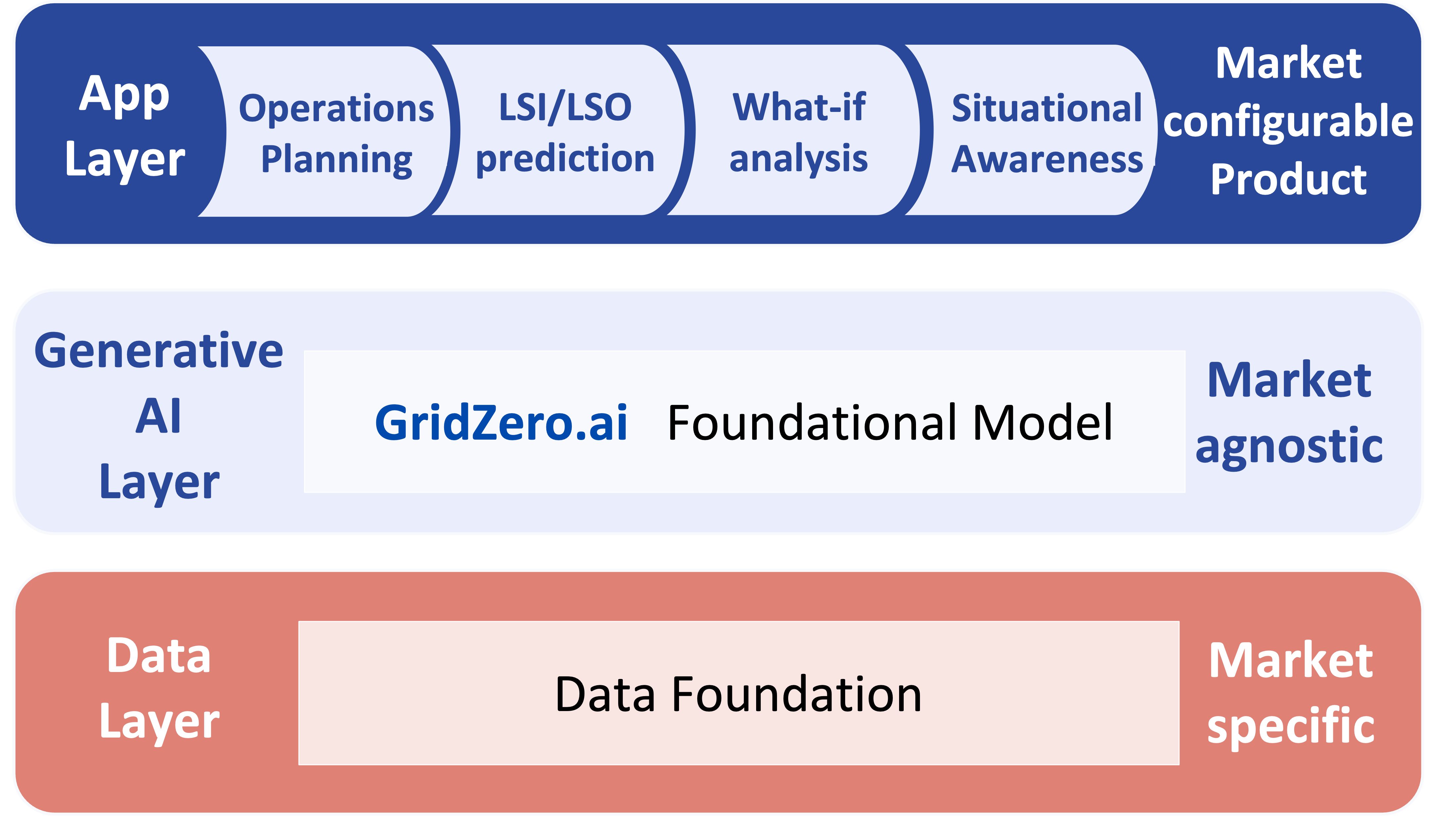}}
    \caption{The GridZero.ai platform consists of three closely connected layers: The unified electricity system and market data layer provides the foundation. Above this, the Gen AI layer learns spatiotemporal relationships that capture the underlying data generation process. Building on these representations, the application layer delivers decision support with finetuned models and customised optimisation loops across a broad range of use cases.}
    \label{fig:gz_tech_stack}
  \end{center}
  \vspace*{-4mm}
\end{figure}

In this section, we describe the proposed Gen AI approach to the LSI/LSO day-ahead forecasting problem.  Specifically, GridZero.ai has developed a purpose-built Gen AI platform to model the electricity system shown in Figure \ref{fig:gz_tech_stack}. It combines the strengths of AI Transformer architectures with domain expertise in electricity systems to encode data efficiently, and to support decision making through inference and optimisation. 
Transformer architectures are superior to conventional machine learning and numerical algorithms at adaptation to real-time context and generalisation\cite{li2024adaptive, kurth2023fourcastnet}. This is particularly valuable for power system forecasting in a rapidly changing system with high levels of uncertainty. 

The GridZero.ai platform provides extensive explainability through built-in access to model internal representations, detailed variable trajectories, scenario-level prediction drivers, and retrieval of comparable historical events. These capabilities allow power system experts to analyze why the model produced a given prediction, identify which signals or system states were influential, and understand how counterfactual changes (e.g., demand, wind, unit outages) propagate through the model. This enables an interactive development approach to key system problems where power system experts can use their knowledge to assess and improve the performance of the solution.

The platform-based solution also enables rapid prototyping and development to inform future production  scale developments.  The end-to-end process to identify and gather required data, train the model, and iterate with subject matter experts (e.g., on the need to model detail of key operational policies, such as the ordering of flows on HVDC interconnectors during network congestion) produced results that took five weeks for the case study presented below.  This is orders of magnitude faster than it would take to configure a traditional solution for the same problem.

%

\section{Case Study}
\label{sec:case}

The case study uses data from August 2022 to August 2024 that covers both the Irish and Great Britain (GB) electricity systems data. The dataset includes both forecast and outturn information relevant to reserve dimensioning, split into training, validation, and test subsets with proportions of 80\%, 10\%, and 10\%, respectively, maintaining temporal order to ensure causal integrity and prevent information leakage.

\begin{itemize}
    \item \underline{\textit{Inputs}}: Forecast features consist of demand, wind, solar generation, and the availability of large generators and interconnectors. Outturn features include the same variables plus curtailment, generator bidding and offer behavior, Ireland-Northern Ireland tie-line (north–south tie-line) exchanges, interconnector exchanges, and the output of the ten largest generators. 
    \item \underline{\textit{Outputs}}: 48 half hourly values of large generator outputs and interconnector flows from D-1 11 pm to D-day 11 pm, along with percentiles around each predicted value, and derived LSI/LSO.
\end{itemize}

The modeling process assumes that the data structure captures the underlying system dynamics through a simplified grid topology and relational data representation with mixed spatial granularity. Temporal and spatial interdependencies across time series are represented, while the model accounts for the fact that only a subset of inputs is available at prediction time. The model is implemented using a Transformer‑style, sequence‑based multi‑input architecture designed to learn spatiotemporal relationships in the data. Temporal dynamics are represented by encoding each signal as a sequence, while spatial correlations arise from jointly modelling dependencies across multiple inputs. The model is trained on operationally relevant features—including wind forecasts, residual load, market price differentials between Ireland and Great Britain, system demand, and generation availability—reflecting the coupled nature of the Irish and GB systems. The attention mechanism enables the model to identify relationships across different time horizons and also supports simple ‘what‑if’ analyses (e.g., assessing how changes in wind forecasts could influence flows, generator behaviour, and the resulting LSI/LSO values).

\subsection{Features that Contribute to LSI and LSO}

\begin{figure}[!h]
  \begin{center}
    \resizebox{\linewidth}{!}{\includegraphics{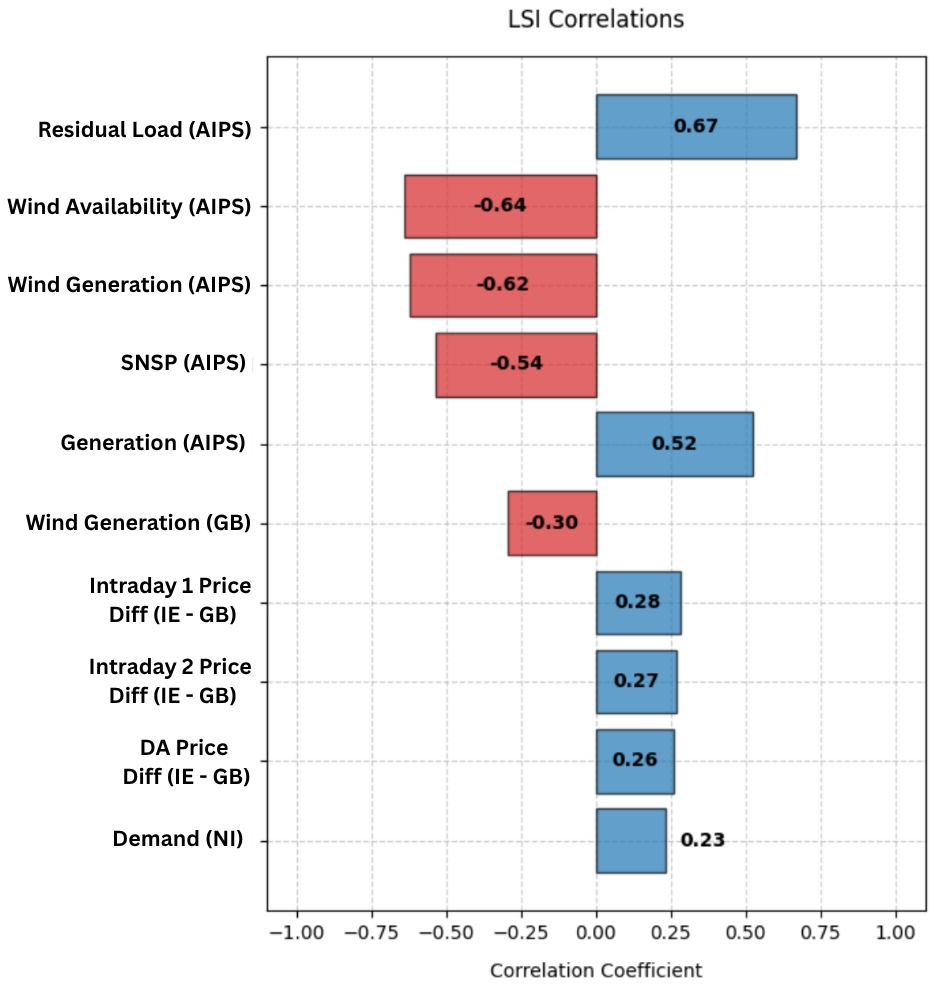}}
    \caption{Pearson correlation coefficients between the LSI value and various system features. Positive values indicate a positive correlation, while negative values represent a negative correlation with the LSI value. (AIPS: All-Island Power System, GB: Great Britain, NI: Northern Ireland, IE: Republic of Ireland).}
    \label{fig:lsi_correlations}
  \end{center}
  \vspace*{-4mm}
\end{figure}

The feature contribution analysis shown in Figure~\ref{fig:lsi_correlations} indicate that LSI value is primarily driven by features such as residual load and renewable generation. High residual load and conventional generation in the AIPS shows strong positive correlations with the LSI, implying that upward reserve requirements increase when demand is high or renewable output is low. Conversely, wind availability and wind generation in AIPS exhibit strong negative correlations, reflecting that higher renewable output tends to reduce LSI values. 

Cross-border dynamics also play a role: positive correlations with intraday and day-ahead price differentials (Ireland–GB) suggest that market conditions and interconnector flows influence reserve requirements. Overall, these correlations highlight that both renewable variability in Ireland \& GB and market conditions jointly shape LSI. The correlation analysis for LSO is not shown, but the market price differential between Ireland and GB has a stronger influence on LSO, as prices largely determine (HVDC) interconnector flow direction.

\subsection{LSI and LSO Results}

Model evaluation follows the same test period across all methods to ensure comparability. Performance metrics include MAPE and precision, recall when compared with actual LSI, LSO values.

To assess realism and performance, the proposed Gen AI forecasts are compared against two established approaches used in system operations:
\begin{itemize}
    \item \underline{\textit{GridZero.ai  developed model}}: The proposed Gen AI model trained on pre-market gate closure system and forecast data available before 9 am D-1. This setup ensures strict temporal causality, with no access to information that becomes available later.
    \item \underline{\textit{Schedule-based D-1 model}}: A simplified approximation of expected LSI and LSO as available at 9 am D-1. It relies on the latest available D-1 expected schedule to form an indicative view of the following day’s system state, serving as a baseline approximation.
    \item \underline{\textit{Rolling 8-hour model}}: A continuous prediction of expected LSI and LSO derived from 8-hour security constrained unit commitment. This benchmark represents an upper bound on forecast accuracy, as it benefits from near-real-time information such as market positions including interconnectors, bids, physical notifications, and updated weather forecasts that are unavailable at the day-ahead stage.
\end{itemize}
Figures 4 and 5 illustrate the LSI and LSO prediction performance of the three models: the GridZero.ai D‑1 model, the schedule‑based D‑1 model, and the rolling 8‑hour benchmark. Each figure compares actual versus predicted values and shows the P5–P95 confidence intervals, highlighting temporal accuracy, variability across approaches, and the relative robustness of the proposed model under limited day‑ahead information. Note that the LSO is more sparse compared to LSI, as it represents only interconnector export events.

\begin{table}[h!]
  \centering
  \caption{LSI and LSO results from the test dataset (June 6, 2024 - August 14, 2024).}
  \label{tab:lsi_lso_results}
  \resizebox{0.95\linewidth}{!}{
  \begin{tabular}{lccc}
    \hline
    \rule{0pt}{2.6ex}\textbf{Method} & \textbf{LSI: MAPE (\%)} & \textbf{LSI: MAE (MW)} & \textbf{LSO: MAE (MW)} \\
    \hline
    \rule{0pt}{2.6ex}GridZero.ai & \multirow{2}{*}{9.8\%} & \multirow{2}{*}{44.5} & \multirow{2}{*}{12.7} \\
    (D-1 9 am) & & & \\
    \hline
    \rule{0pt}{2.6ex}Schedule-based & \multirow{2}{*}{19.9\%} & \multirow{2}{*}{89.7} & \multirow{2}{*}{21.9} \\
    (D-1 9 am) & & & \\
    \hline
    \rule{0pt}{2.6ex}Schedule-based & \multirow{2}{*}{8.7\%} & \multirow{2}{*}{39.3} & \multirow{2}{*}{6.1} \\
    (Rolling Hour-8) & & & \\
    \hline
  \end{tabular}}
\end{table}

\begin{figure}[!h]
  \begin{center}
    \resizebox{\linewidth}{!}{\includegraphics{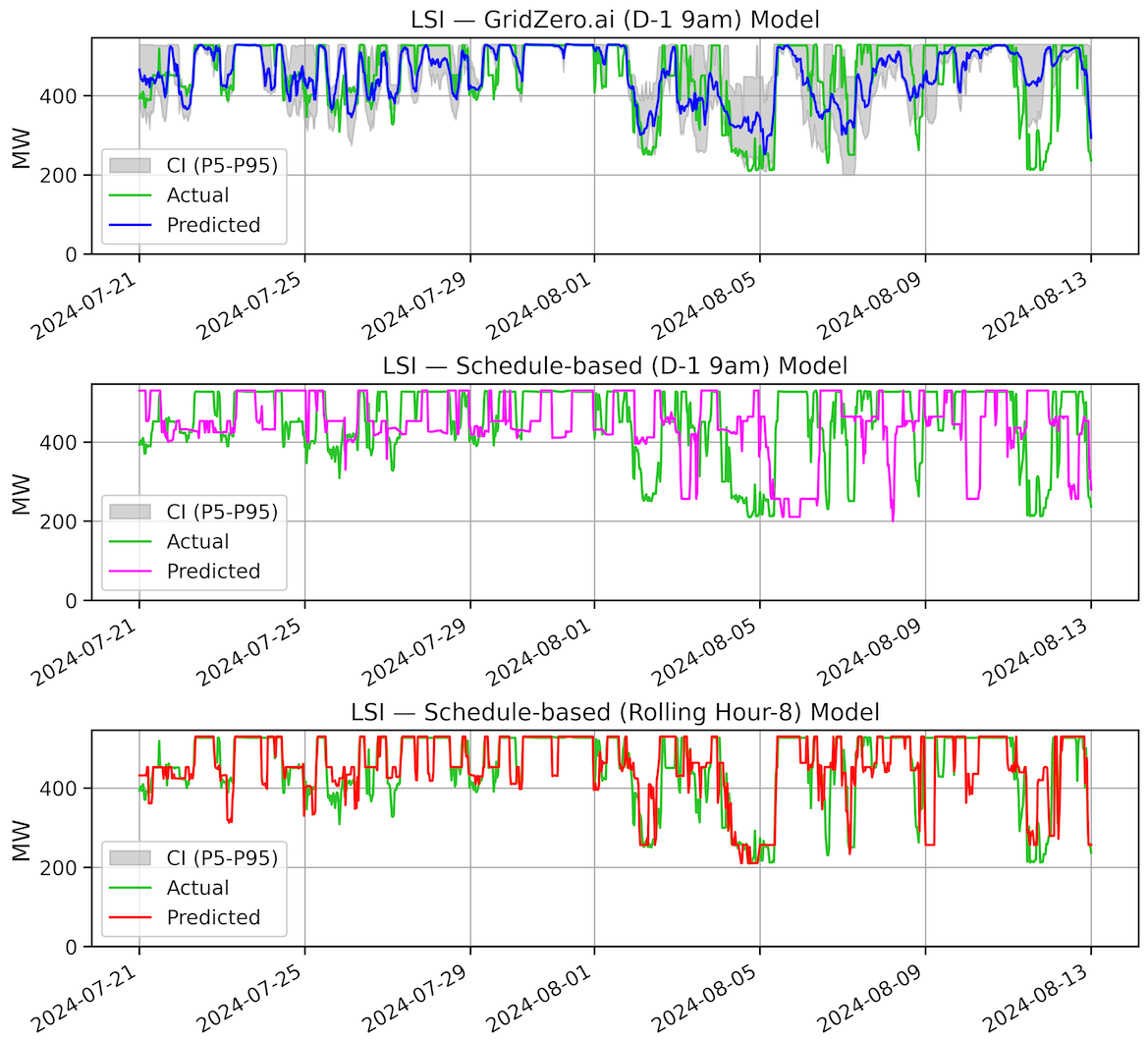}}
    \caption{Actual vs. predicted LSI values for the three models—(top) GridZero.ai D‑1 9 am model, (middle) schedule‑based D‑1 9 am model, and (bottom) rolling 8‑hour model—with P5–P95 confidence intervals.}
    \label{fig:lsi_ts_results}
  \end{center}
  \vspace*{-4mm}
\end{figure}
\begin{figure}[!h]
  \begin{center}
    \resizebox{\linewidth}{!}{\includegraphics{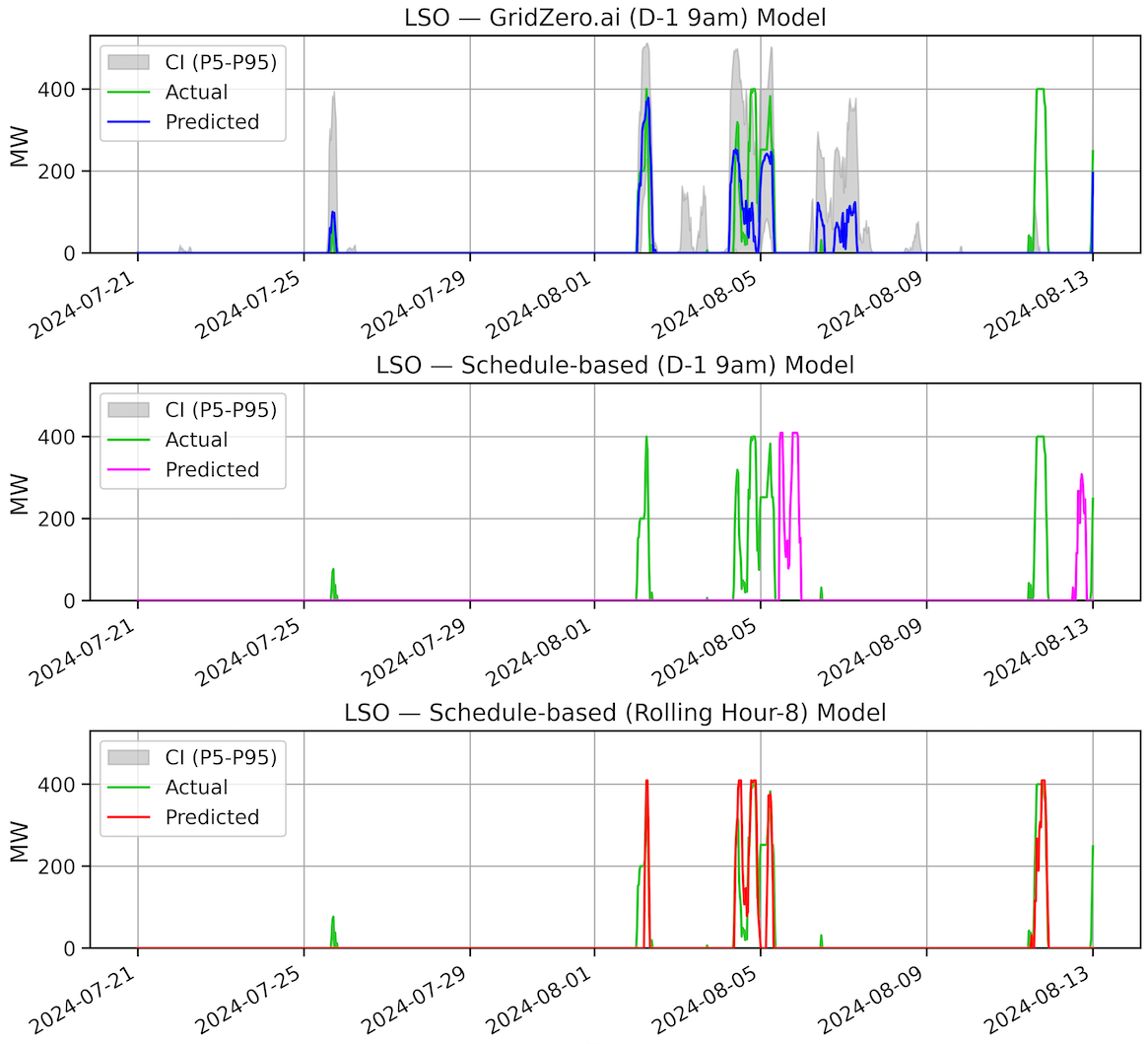}}
    \caption{Actual vs. predicted LSO values for the three models—(top) GridZero.ai D‑1 9 am model, (middle) schedule‑based D‑1 9 am model, and (bottom) rolling 8‑hour model—with P5–P95 confidence intervals.}
    \label{fig:lso_ts_results}
  \end{center}
  \vspace*{-4mm}
\end{figure}

Across the evaluation period, the GridZero.ai developed model achieves a MAPE that is only 1.1\% higher than the operational 8-hour rolling model despite operating with substantially less information and a longer lead time. This operational benchmark reflects the system operator’s current practical capability for reserve‑related forecasting prior to the introduction of DASSA, rather than a theoretical forecasting limit. The comparison is conducted over identical evaluation periods and with aligned input data to ensure a fair and consistent assessment. Feature contribution analysis indicates that interconnector flows, wind generation forecasts, and large-unit availability are the most influential predictors of LSI and LSO variations. Sensitivity tests suggest that early and accurate forecasts of these variables could translate to around 15\% potential reduction in reserve costs when utilizing the P95 predictions as opposed to a standard flat-line reserve procurement. The results confirm the feasibility of applying Gen AI to forecast key security indicators in advance of market closure, providing actionable insights for day-ahead reserve planning and broader operational decision-making. 

%
\section{Conclusions}
\label{sec:conclu}

This paper presents a Gen AI approach for forecasting LSI and LSO in a real-world low-inertia power system to support reserve dimensioning.  Developed collaboratively by EirGrid and GridZero.ai, the system produced accurate day-ahead forecasts up to 38 hours ahead using only pre-market information.  The AI-based method achieved results with a MAPE that is only 1.1\% higher than the operational 8-hour model, indicating the potential for a significant reduction in reserve capacity procurement costs while maintaining operational realism.  

Future work will focus on possible production integration for AI-informed reserve procurement and decision support, with further extension toward real-time operational applications as power systems become increasingly dynamic and data-driven. This evaluation is based on a five‑week test window with comparisons against an operational solver. While this demonstrates initial performance under real conditions, a longer evaluation period—including more atypical system states—would provide a broader assessment. Future work will also include comparisons with additional ML forecasting baselines and more detailed robustness and out‑of‑distribution analyses to further evaluate generalisation capability.

\bibliographystyle{IEEEtran}
\bibliography{refs}

\end{document}